\documentclass[aps,12pt,superscriptaddress,amsfonts,amssymb,amsmath]{revtex4}

\usepackage{graphicx}
\usepackage{epsfig}
\usepackage{makeidx}
\usepackage{epstopdf}

\begin{document}

\title{Electromagnetic coupling of strongly non-local \\ quantum mechanics}

\author{G.\ Modanese \footnote{Email address: giovanni.modanese@unibz.it}}
\affiliation{Free University of Bolzano-Bozen \\ Faculty of Science and Technology \\ I-39100 Bolzano, Italy}

\linespread{0.9}

\begin{abstract}

\bigskip

Although standard quantum mechanics has some non-local features, the probability current of the Schr\"odinger equation is locally conserved, and this allows minimal electromagnetic coupling. For some important extensions of the Schr\"odinger equation, however, the probability current is not locally conserved. We show that in these cases the correct electromagnetic coupling requires a relatively simple extension of Maxwell theory which has been known for some time and recently improved by covariant integration of a scalar degree of freedom. We discuss some general properties of the solutions and examine in particular the case of an oscillating dipolar source. Remarkable mathematical and physical differences emerge with respect to Maxwell theory, as a consequence of additional current terms present in the equations for $\nabla \cdot \textbf{E}$ and $\nabla \times \textbf{B}$. Several possible applications are mentioned.

\end{abstract}

\maketitle

\section{Introduction. Non-local wave equations}

It is well known that quantum mechanics has some non-local features, as seen for instance in the EPR phenomenon and in the possibility of state teleportation. Such features are well understood and are actually being exploited in a new generation of quantum computing devices. Still the probability current of the Schr\"odinger equation obeys a strictly local conservation rule, or ``continuity equation'', namely
\begin{align}
	\frac{{\partial \rho }}{{\partial t}} + \nabla  \cdot {\bf{J}} = 0; \qquad \rho  = |\Psi {|^2}; \qquad {\bf{J}} = \frac{{ - i\hbar }}{{2m}}\left( {{\Psi ^*}\nabla \Psi  - \Psi \nabla {\Psi ^*}} \right) .
	\label{cont}
\end{align}

This is also true in the case of many-particle systems described in second quantization formalism with local interactions, and for extensions of the Schr\"odinger equation which describe macroscopic quantization, like the Ginzburg-Landau equation. In all these cases the electromagnetic coupling of the system is readily obtained via gauge invariance, by replacing the momentum with $\left( {{\bf{p}} - q{\bf{A}}/c} \right)$ and the energy by $\left( {E - qV} \right)$; this is in turn equivalent to the coupling ${J_\mu }{A^\mu }$, where ${J^\mu }$ and ${A^\mu }$ are the four-vectors $J^\mu=(c\rho,\textbf{J})$, $A^\mu=(V,\textbf{A})$. Classically, the field ${A^\mu }$ satisfies the Maxwell equations $\partial_\mu F^{\mu \nu}=4\pi J^\nu/c$, and since ${F^{\mu \nu }}$ is antisymmetric these are only compatible with four-currents which are locally conserved, i.e.\ obey the continuity equation (\ref{cont}), which can also be written as $\partial_\mu J^\mu=0$.

There exist, however, important extensions of the Schr\"odinger equation which do not admit in general a conserved current. A well-known example is the equation
\begin{align}
	i\hbar\frac{\partial}{\partial t}\Psi(x,t)=-\frac{\hbar^2}{2m} \frac{\partial^2}{\partial x^2} \Psi(x,t) + \int_0^t d\tau \int_{-\infty}^{+\infty} dy \, U(x-y,t-\tau) \Psi(y,\tau),
\end{align}
which involves the non-local interaction kernel $U$ (possibly with memory) and whose solutions have been studied by Lenzi et al.\ \cite{Lenzi2008solutions}. Such equations have been employed for the effective description of systems with long range interactions \cite{latora1999superdiffusion}, enhanced diffusion in active intracellular transport \cite{caspi2000enhanced} and nuclear scattering \cite{chamon1997nonlocal,balantekin1998green}. An  earlier example of a non-local wave equation is the Gorkov equation for the superconducting order parameter $\Delta (\textbf{x})$, which takes the form $\Delta (\textbf{x})=\int d^3 y \, K(\textbf{x},\textbf{y}) \Delta (\textbf{y})$ and can be approximated by a Ginzburg-Landau equation in suitable limits \cite{waldram1996superconductivity}.

Furthermore, it has been recently recognized that fractional quantum mechanics also lacks a locally conserved current. Starting from the fractional Schr\"odinger equation \cite{laskin2002fractional}, in which the kinetic energy operator is proportional to $|\textbf{p}|^\alpha$, $1<\alpha \le 2$, Wei \cite{wei2016comment} has pointed out that the correct continuity equation involves an extra source term, namely 
\begin{align}
	\frac{{\partial \rho }}{{\partial t}} + \nabla  \cdot {\bf{J}} = I_\alpha  = -i \hbar^{\alpha-1} D_\alpha \left[ \nabla {\Psi ^*}(-\nabla^2)^{\frac{\alpha}{2}-1}\nabla \Psi  - c.c. \right] .
	\label{weiI}
\end{align}
The applications of the fractional Schr\"odinger equation are manifold \cite{Lenzi2008fractional}, and therefore this property is important. Its conceptual meaning is intriguing. For instance, numerical solutions in the form of discrete or continuous wave packets display an explicit phenomenon of back-diffusion with $I_\alpha \neq 0$ and locally non-conserved probability (Sect.\ \ref{wav}) \cite{Modanese2017480}.

For all these systems the electromagnetic coupling cannot take the usual form, with local gauge invariance and locally conserved current. A suitable extension of the Maxwell equations is needed, and such an extension actually exists, called Aharonov-Bohm electrodynamics. This has been studied by several authors (\cite{hively2012toward,van2001generalisation} and refs.) and in \cite{Modanese2017MPLB} a compact covariant version has been found, in which the degree of freedom $S=\partial_\mu A^\mu$ is explicitly eliminated. Here (Sect.\ \ref{ext}) we first recall  briefly this covariant form, that is important because it shows that despite all the non-locality involved and its possible startling consequences, the theory is fully compatible with Special Relativity. Then we derive the modified Maxwell equations in the familiar 3D vector form with Gauss units, in order to evidence the modifications introduced with respect to Maxwell’s theory, through additional current terms whose physical effects, we believe, can be tested and can lead to potentially useful applications.

In Sect.\ \ref{osc} we write the solution for a stylized oscillating dipole. In Sect.\ \ref{wav} we compute explicitly the extra-source term (\ref{weiI}) for a wave packet. Sect.\ \ref{con} contains our conclusions.

\section{Extended Maxwell equations}
\label{ext}

The extended Maxwell equations in covariant form \cite{Modanese2017MPLB} are obtained from the Aharonov-Bohm Lagrangian, by elimination of the scalar degree of freedom $S$. In Gauss-CGS units they read
\begin{align}
	\partial_\mu F^{\mu \nu}=\frac{4\pi}{c}J^\nu-\frac{4\pi}{c}\partial^\nu\partial^{-2} \left(\partial_\gamma J^\gamma \right).
	\label{mme-cov}
\end{align}
We use standard notations and conventions as for instance in Jackson \cite{jackson1975electrodynamics}. All quantities are functions of $(t,\textbf{x})$. The equations without sources coincide with the corresponding Maxwell equations, namely $\partial^\rho \varepsilon_{\sigma \rho \mu \nu} F^{\mu \nu}=0$. We shall denote by $I$ the four-divergence of the current:
\begin{align}
	I=\left(\partial_\gamma J^\gamma \right)=\frac{\partial\rho}{\partial t}+\nabla \cdot \bf{J},
\end{align}
where $\rho$ and $\textbf{J}$ are computed from the wave function of the system under consideration. The equations (\ref{mme-cov}) represent essentially the only possible covariant extension of the Maxwell theory, in which a residual gauge invariance is present. The inverse operator of the D'Alembertian, namely $\partial^{-2}=\Box^{-1}$, with appropriate causal boundary conditions, can be expressed through a retarded integral; its action on a function $f(t,\textbf{x})$ is as follows:
\begin{align}
	& \partial^2 g(t,\textbf{x})=f(t,\textbf{x})\Rightarrow g(t,\textbf{x})= \partial^{-2} f(t,\textbf{x})\Rightarrow\\ \nonumber 
	& \Rightarrow g(t,\textbf{x})=\frac{1}{4\pi} \int d^3y \frac{f\left(t-\frac{1}{c}|\textbf{x}-\textbf{y}|,\textbf{y} \right)}{|\textbf{x}-\textbf{y}|} \equiv \frac{1}{4\pi} \int d^3y \frac{f\left(t_{ret},\textbf{y} \right)}{|\textbf{x}-\textbf{y}|}.
\end{align}
The operator $\partial^{-2}$ is non-local.
This is the origin of an important feature of these equations, namely that the effective source of the fields is the sum of the ``primary'' physical and localized source $J^\nu$, and a ``secondary'' source which extends outside the region where the wave function is not zero. This will emerge clearly in the case of the oscillating dipolar source (Sect.\ \ref{osc}).

In the familiar 3D vector formalism the extended equations without sources are written as usual, namely $\nabla \times \textbf{E}=-(1/c)(\partial \textbf{B}/\partial t)$, $\nabla \cdot \textbf{B}=0$. The extended equations with sources take the form
\begin{align}
	& \nabla \cdot \textbf{E}=4\pi \rho-\frac{1}{c^2}\frac{\partial}{\partial t}\int d^3y \frac{I\left(t_{ret},\textbf{y} \right)}{\left|\textbf{x}-\textbf{y} \right|}; \label{eqE} \\
	& \nabla \times \textbf{B}-\frac{1}{c} \frac{\partial \textbf{E}}{\partial t}=\frac{4\pi}{c} \textbf{J}+\frac{1}{c} \nabla \int d^3y \frac{I\left(t_{ret},\textbf{y} \right)}{\left|\textbf{x}-\textbf{y} \right|}. \label{eqB}
\end{align}

Although the theory is not gauge invariant, at the mathematical level these equations can still be solved by introducing auxiliary potentials (thanks to the equations for $\nabla \times \textbf{E}$ and $\nabla \cdot \textbf{B}$ and to uniqueness theorems \cite{woodside2009three}). We shall denote by $\phi_{aux}$ and $\textbf{A}_{aux}$ auxiliary potentials in the Feynman-Lorenz gauge. The auxiliary potentials can be used as intermediate functions for the solutions of the field equations, but they cannot be used in a Hamiltonian formulation of the theory, which in principle is possible but must employ physical potentials fixed through a very complex residual gauge condition \cite{Modanese2017MPLB}. This is the price that must be paid for the loss of gauge invariance and local current conservation in the theory. For this reason (see also our conclusions in Sect.\ \ref{con}) we shall consider here the electromagnetic field in a semi-classical approximation, i.e., only its source $\Psi$ is quantized.

The equations for the auxiliary potentials are
\begin{align}
	& \frac{1}{c^2}\frac{\partial^2 \phi_{aux}}{\partial t^2}-\nabla^2 \phi_{aux}=4\pi \rho-\frac{1}{c^2} \frac{\partial}{\partial t} \int d^3y \frac{I\left(t_{ret},\textbf{y} \right)}{\left|\textbf{x}-\textbf{y} \right|}; \label{pot1} \\ 
	& \frac{1}{c^2}\frac{\partial^2 \textbf{A}_{aux}}{\partial t^2}-\nabla^2 \textbf{A}_{aux}=\frac{4\pi}{c} \textbf{J}+\frac{1}{c} \nabla \int d^3y \frac{I\left(t_{ret},\textbf{y} \right)}{\left|\textbf{x}-\textbf{y} \right|}. \label{pot2}
\end{align}
The relation with the fields is of course, as usual, $\textbf{E}=-c^{-1}\partial_t \textbf{A}_{aux}-\nabla \phi_{aux}$, $\textbf{B}=\nabla \times \textbf{A}_{aux}$. 

\section{Oscillating dipolar source}
\label{osc}

We now study the solution of the extended equations for a stylized source which is physically relevant and formally manageable. In the equations (\ref{pot1}), (\ref{pot2}) for the potentials the mathematical difficulty is evident: a double retarded integration is required, since $\phi_{aux}$ and $\textbf{A}_{aux}$ are in turn retarded integrals of the r.h.s. So one can proceed if at least the first integration can be done analytically. We have discussed in \cite{Modanese2017MPLB} the case of the static magnetic field generated by a current tunnelling between infinite parallel planes. Here we consider an oscillating dipole with a non-locally-conserved current of the form
\begin{align}
	\rho & =q {\mathop{\rm Re}\nolimits} \left\{ e^{i \omega t} \left[\delta^3(\textbf{x}-\textbf{a})-\delta^3(\textbf{x}+\textbf{a}) \right] \right\}; \label{j0}\\
	\textbf{J} & =0. \label{jj}
\end{align}
This describes two opposite point-like charges placed at $\textbf{x}=\pm \textbf{a}$, which oscillate between values $\pm q$ with frequency $\omega$, without an intermediate current. More generally, the oscillating dipolar source can have a current which is insufficient to account for the oscillation of the charges (compare Sect.\ \ref{wav}). So we can regard the expression above as an extreme example of a non-conserved current, which in real cases may be only one of the components of a real source.

Let us compute the four-divergence $I={\partial_\mu j^\mu }$. The only contribution is given by the time derivative of ${j^0}$:
\begin{align}
	I = \frac{\partial \rho}{\partial t}  = \frac{1}{2}iq\omega \left( {{e^{i\omega t}} - {e^{ - i\omega t}}} \right)\left[ {{\delta ^3}\left( {{\bf{x}} - \textbf{a}} \right) - {\delta ^3}\left( {{\bf{x}} + \textbf{a}} \right)}. \right]
\end{align}
The retarded integral of $I$, to be inserted into the extended equations for the fields (\ref{eqE}), (\ref{eqB}) or into those for the potentials (\ref{pot1}), (\ref{pot2}) is
\begin{align}
	\int d^3y \, \frac{I(t_{ret},\textbf{y})}{|\textbf{x}-\textbf{y}|} =
	\frac{1}{2}iq\omega \left[ \frac{e^{i\omega \left(t-\frac{1}{c}|\textbf{x}-\textbf{a}| \right)}}{|\textbf{x}-\textbf{a}|} - \frac{e^{i\omega \left(t-\frac{1}{c}|\textbf{x}+\textbf{a}| \right)}}{|\textbf{x}+\textbf{a}|} - c.c. \right].
\end{align}
From this one can write an expression for the auxiliary potentials. For instance one obtains for $\phi_{aux}$, after another retarded integration
\begin{align}
	\phi_{aux}(t,\textbf{x})=& \frac{1}{2} q\left[ \frac{e^{i\omega \left(t-\frac{1}{c}|\textbf{x}-\textbf{a}| \right)}}{|\textbf{x}-\textbf{a}|} - \frac{e^{i\omega \left(t-\frac{1}{c}|\textbf{x}+\textbf{a}| \right)}}{|\textbf{x}+\textbf{a}|} + c.c. \right] + \\ \nonumber
	& - \frac{1}{4\pi c^2} \frac{\partial}{\partial t}\int d^3y \, \frac{\frac{1}{2}iq\omega}{|\textbf{x}-\textbf{y}|} 
	 \left[ \frac{e^{i\omega \left(t-\frac{1}{c}|\textbf{y}-\textbf{a}| \right)}}{|\textbf{y}-\textbf{a}|} - \frac{e^{i\omega \left(t-\frac{1}{c}|\textbf{y}+\textbf{a}| \right)}}{|\textbf{y}+\textbf{a}|} - c.c. \right].
\end{align}
In a separate work we shall give a numerical evaluation of the fields obtained from these potentials. Many of the familiar approximation schemes clearly cannot be applied, since the secondary source is quite unusual, extending over all space.

In order to study in analytical form at least one simplified case, let us rewrite eq.\ (\ref{eqE}) as 
\begin{align}
	\nabla \cdot \textbf{E}=4\pi \rho + 4\pi \rho',
\end{align}
where
\begin{align}
	\rho'=-\frac{1}{4\pi c^2} \frac{\partial}{\partial t}\int d^3y \frac{I\left(t_{ret},\textbf{y} \right)}{\left|\textbf{x}-\textbf{y} \right|}.
\end{align}
Now consider the electric dipole moments of the densities $\rho$ and $\rho'$. In the low frequency and near field limits, the electric field will be essentially determined by these two dipole moments, since the magnetic contributions are smaller by many orders of magnitude. (This can be checked by including the appropriate transport current in (\ref{jj}), such that the continuity equation holds, and then solving the standard Maxwell equations.) The dipole moment ${\bf d}$ of $\rho$ is readily found: $\textbf{d}=q\textbf{a}\cos(\omega t)$. The moment of $\rho'$ is obtained through a long but straightforward integration, whose result is
\begin{align}
	\textbf{d}'& =\int d^3x \, \textbf{x} \rho'(\textbf{x}) = 
	 q\textbf{a} \left[ \frac{\omega R}{c} \sin(kR-\omega t)+\cos(kR-\omega t)-\cos(\omega t) \right].
\end{align}
Here we have integrated up to the distance $R$ from the origin and we suppose that $R \ll \lambda$, where $k=\omega/c=\lambda^{-1}$. Let us assume that the contribution of the secondary charge to the electric field at distance $R$ is given by its dipole moment integrated up to $R$. This is an heuristic assumption and still does not really give us the field (because we must do a further integration), but it is interesting because (a) it illustrates the role of the secondary spatial charge in the extended Maxwell equations, and (b) it shows that its contribution is comparable to that of the primary charge. In fact, the third term in $\textbf{d}'$ cancels the contribution of $\rho$; the first term in $\textbf{d}'$ is much smaller than the others, in near field, because it is proportional to $R/\lambda$; and the second term reproduces almost exactly the dipole moment of $\rho$, except for the small phase shift $kR$. 

All this confirms the ``censorship'' mechanism discussed in \cite{Modanese2017MPLB}, according to which measurements of the field strength made with test particles cannot reveal the local non-conservation of the source. But this illusion of conservation is due to the secondary source, which in fact is extended in space. We expect this to have consequences (still to be explored in detail), for instances on cases where there are conductors near the source imposing boundary conditions on the field.

\section{Wave packet in fractional quantum mechanics}
\label{wav}

\begin{figure}
\begin{center}
\includegraphics[width=10cm,height=7cm]{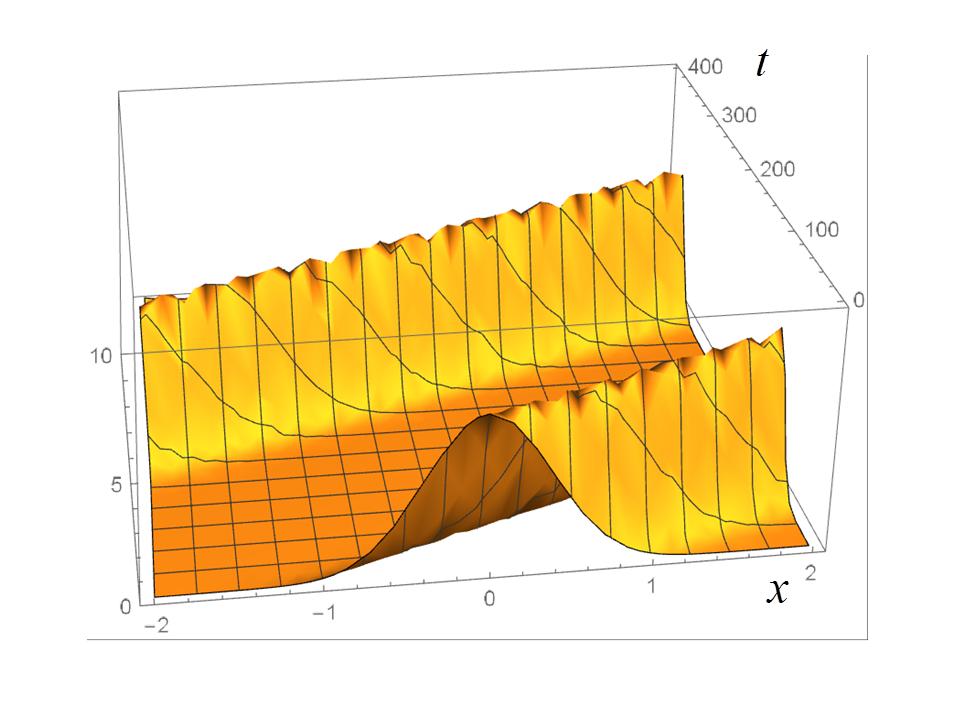}
\caption{Squared amplitude of a discrete wave packet in fractional quantum mechanics ($\alpha=3/2$) made of a sum of 20 plane waves with positive momentum. The ``main'' branch of the packet, which has its maximum at $x=0$ for $t=0$, is evidenced. Another branch with maximum at negative $x$ for $t=0$ is visible behind it.} 
\label{fig1}
\end{center}  
\end{figure}

\begin{figure}
\begin{center}
\includegraphics[width=10cm,height=7cm]{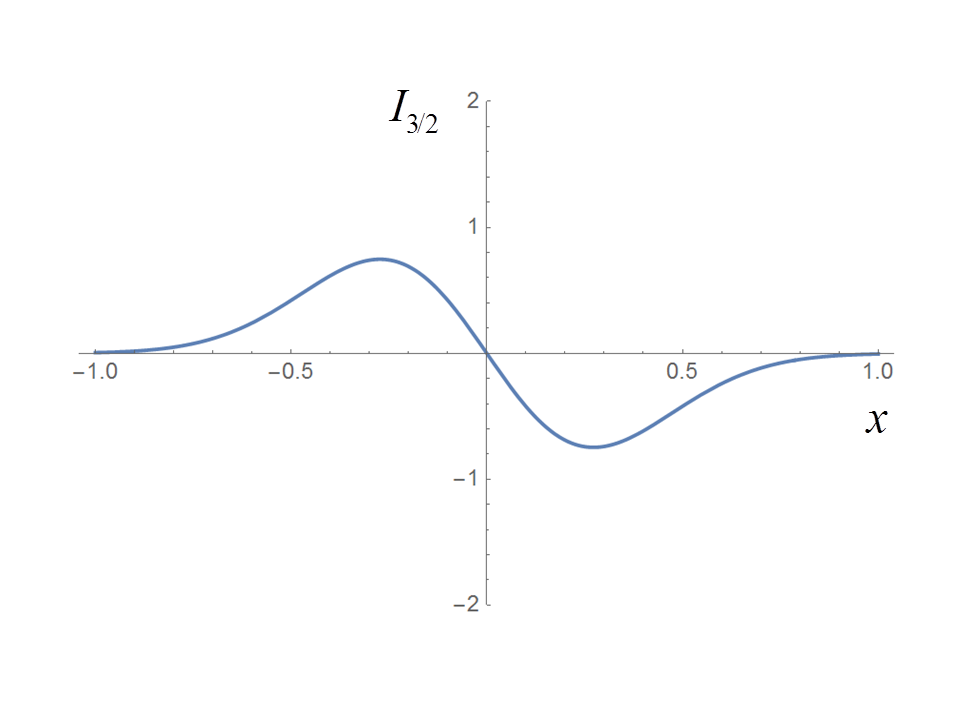}
\caption{Extra-source term at $t=0$ for the main branch of the packet in Fig.\ \ref{fig1}, showing sub-diffusion.} 
\label{fig2}
\end{center}  
\end{figure}

In this section we compute the extra-source term $I_\alpha(t,\textbf{x})$ for a wave packet in fractional quantum mechanics and find that it contains regions with positive and negative balance. This provides a motivation for the definition of the dipolar source of the previous section and shows explicitly how sub-diffusion and local probability non-conservation occur in fractional quantum mechanics in this case \cite{tayurskii2012superfluid}. Starting from the expression (\ref{weiI}) for the extra-source term, Wei \cite{wei2016comment} has computed $I_\alpha$ for a superposition of two plane waves. There are in the literature no expressions for $I_\alpha$ in other cases, although for instance Lenzi et al.\ have found several explicit solutions of their wave equation \cite{Lenzi2008solutions}, which is closely related to a fractional Schr\"odinger equation. In \cite{Modanese2017480} we have computed $I_{3/2}$ for a non-normalizable discrete packet, given by a finite sum of $n$ plane waves. In that case $I_{3/2}$ can be written analytically and a plot is given in Fig.\ \ref{fig2}, for a case with $n=20$. It is also possible to plot the time dependence of $|\Psi|$, which shows that the packet moves without spreading (Fig.\ \ref{fig1}). In view of the form of $I_{3/2}$, one can interpret the sub-diffusive movement of the packet as involving a steady backward internal displacement of probability without a corresponding current (there is a net destruction of probability on the right of the maximum and a net creation on the left, when the packet moves to the right).

Similar results can be obtained with a continuous packet of the form
\begin{align}
	\Psi(t,\textbf{x})=\int d^3p \, e^{i\hbar^{-1}\textbf{p}\textbf{x}} \tilde{\Psi}(t,\textbf{p}).
\end{align}
The operator $(-\nabla^2)^{\frac{\alpha}{2}-1}$ acts on $\tilde{\Psi}$ multiplying it by $|\textbf{p}|^{\alpha-2}$. So we obtain for $I_\alpha$ the integral
\begin{align}
	I_\alpha(t,\textbf{x})=-i\hbar^{\alpha-1} D_\alpha \int d^3p \, \int d^3q \, \left[ e^{i\hbar^{-1}(\textbf{p}-\textbf{q})\textbf{x}} (\textbf{p}\textbf{q}) |\textbf{p}|^{\alpha-2} \tilde{\Psi}(t,\textbf{p}) \tilde{\Psi}^*(t,\textbf{q}) - c.c. \right].
	\label{intI}
\end{align}
This integral can be computed numerically, with an appropriate ansatz for $\tilde{\Psi}$. Details will be given elsewhere.

\section{Conclusions}
\label{con}

We have obtained an essentially unique extension of the Maxwell equations which allows to couple the classical electromagnetic field to a microscopic or macroscopic quantum system whose wave function does not have a locally conserved probability current. We have written the new equations, which we had originally derived in covariant 4D form and in natural-Heaviside units, in 3D vector form and in CGS-Gauss units. This allows to spot clearly the additional terms with respect to the usual Maxwell equations, namely a secondary non-local charge density in the equation for $\nabla \cdot \textbf{E}$, and a secondary non-local current in the equation for $\nabla \times \textbf{B}$. We have mentioned some physical systems where the probability current may be not locally conserved, writing explicitly the extra-source $I=\frac{\partial \rho}{\partial t} + \nabla \cdot \textbf{J}$ for a wave packet in fractional quantum mechanics. We have written the solution of the extended Maxwell equations in integral form for an extra-source with a double $\delta$-function which can represent the limit of a very localized wave packet with oscillating dipolar charge. A detailed numerical evaluation of the solution will be given elsewhere. We have examined heuristically the low-frequency limit by computing the electric dipole moment of the secondary charge amd comparing it with the dipole of the primary charge. The general conclusion is that the extended equations contain new physics, and not only very small corrections, in those cases where the quantum source effectively has an extra-source term. We expect this to be relevant especially for wave functions describing condensed matter systems with a macroscopic number of particles, either with phase coherence (like for superconductors) or not (like for proximity effects in semiconductors), which can be modeled through a non-local, fractional or Gorkov equation.

Further possible developments concern phenomenological models of nuclear interactions, which often employ extended Schr\"odinger equations with a non-local term \cite{chamon1997nonlocal,balantekin1998green}, or with energy-dependent potential terms, which are also  not consistent with the continuity equation for probability \cite{Formanek2004,Lombard2007}. Since in this case the systems are microscopic, their electromagnetic emission is probably not relevant. However, the Coulomb interaction between the nuclei could be influenced by the form of the wave function, with its internal non-conservative probability transport (see Sect.\ \ref{wav} for sub-diffusion in fractional quantum mechanics; in other cases super-diffusion may be present). This possibility deserves further attention.

\bibliographystyle{unsrt}
\bibliography{mme}
 
\end{document}